\documentclass[article]{jss}

\usepackage{orcidlink,thumbpdf,lmodern}
\usepackage{framed}
\usepackage{amsmath}
\usepackage{bm}
\usepackage{bbm}
\usepackage{hyperref}
\usepackage{natbib}
\usepackage{color}
\usepackage{float}
\usepackage{fancyvrb}
\usepackage{comment}

\newcommand{\bfalpha}{\boldsymbol{\alpha}}
\newcommand{\bfdelta}{\boldsymbol{\delta}}
\newcommand{\bfmu}{\boldsymbol{\mu}}
\newcommand{\bfvartheta}{\boldsymbol{\vartheta}}
\newcommand{\bfy}{\boldsymbol{y}}

\author{Skylar Shi~\orcidlink{0009-0001-2818-0299}\\University of Washington
   \And Abel Rodriguez~\orcidlink{0000-0001-5503-7394}\\University of Washington
   \And Rayleigh Lei~\orcidlink{0000-0002-0444-9708}\\University of Michigan}
\Plainauthor{Skylar Shi, Abel Rodriguez, Rayleigh Lei}

\title{\pkg{pumBayes}: Bayesian Estimation of Probit Unfolding Models for Binary Preference Data in \proglang{R}}
\Plaintitle{pumBayes: Bayesian Estimation of Probit Unfolding Models for Binary Preference Data in R}
\Shorttitle{\pkg{pumBayes}: Probit Unfolding Models in \proglang{R}}

\Abstract{Probit unfolding models (PUMs) are a novel class of scaling models that allow for items with both monotonic and non-monotonic response functions and have shown great promise in the estimation of preferences from voting data in various deliberative bodies.  This paper presents the \proglang{R} package \pkg{pumBayes}, which enables Bayesian inference for both static and dynamic PUMs using Markov chain Monte Carlo algorithms that require minimal or no tuning. In addition to functions that carry out the sampling from the posterior distribution of the models, the package also includes various support functions that can be used to pre-process data, select hyperparameters, summarize output, and compute metrics of model fit.  We demonstrate the use of the package through an analysis of two datasets, one corresponding to roll-call voting data from the 116th U.S.\ House of Representatives, and a second one corresponding to voting records in the U.S.\ Supreme Court between 1937 and 2021.}

\Plainkeywords{unfolding models, spatial voting models, non-monotonic response function, item response theory, Markov chain Monte Carlo, R, C++}

\Address{
  Skylar Shi\\
  Graduate Student\\
  Department of Statistics\\
  University of Washington\\
  1410 NE Campus Pkwy\\
  Seattle, WA, USA\\
  E-mail: \email{dshi98@uw.edu}
}

\begin{document}
% \input{article-concordance}

%% -- Introduction -------------------------------------------------------------

%% - In principle "as usual".
%% - But should typically have some discussion of both _software_ and _methods_.
%% - Use \proglang{}, \pkg{}, and \code{} markup throughout the manuscript.
%% - If such markup is in (sub)section titles, a plain text version has to be
%%   added as well.
%% - All software mentioned should be properly \cite-d.
%% - All abbreviations should be introduced.
%% - Unless the expansions of abbreviations are proper names (like "Journal
%%   of Statistical Software" above) they should be in sentence case (like
%%   "generalized linear models" below).

\section{Introduction} \label{sec:intro}

Scaling, ``the construction of measures by associating qualitative judgments about unobservable constructs with quantitative, measurable metric units'' \citep{bhattacherjee2012social}, remains one of the most common but difficult tasks in empirical social science research.  Many modern scaling techniques rely on variations and extensions of statistical approaches that combine factor analysis and generalized linear models.  However, unlike other applications of factor models, the key quantity of interest in scaling problems are the estimates of the underlying latent traits associated with each individual/experimental unit.

In the case of instruments composed exclusively of binary items, Item Response Theory (IRT) models \citep{fox2010bayesian} are well established scaling tools, particularly in the psychometric literature. Correspondingly, software that can fit IRT models to data is widely available.  In the case of the \proglang{R} computing environment, examples of packages that can accomplish this include \pkg{eRM} \citep{mair2007extended}, \pkg{ltm} \citep{rizopoulos2006ltm}, \pkg{mirt} \citep{chalmers2012mirt}, \pkg{psychotree} \citep{strobl2015rasch,komboz2018tree} and \pkg{MCMCpack} \citep{Martin2011MCMCpack}.  Furthermore, IRT models can also be fitted using general purpose multilevel modeling packages such as \pkg{lme4} \citep{bates2014fitting}, \pkg{lavaan} \citep{rosseel2012lavaan}, \pkg{MCMCglmm} \citep{hadfield2010mcmc} and \pkg{bmrs} \citep{burkner2017brms,burkner2019bayesian}.  Here, we are particularly interested in scaling applications in political science, and specially those related to the estimation of revealed preferences from voting data in deliberative bodies such as the U.S.\ Congress or the U.S.\ Supreme Court, with the latent scale interpreted as representing the ``ideology'' of the legislator or judge in a liberal-conservative scale.  In that context, IRT models can be seen as examples of spatial voting models \citep{davis1970expository,Enelow1984}.  Various \proglang{R} packages for fitting models in this class are available, including \pkg{wnominate} \citep{poole2024wnominate} and \pkg{pscl} \citep{jackman2024pscl,zelleis2008regression}.

Most of the latent variable models for binary data implemented in the packages listed above implicitly assume that the response functions of the items (which describes how the probability of a positive outcomes depends on the latent trait) are monotonic.  While such an assumption is appropriate in many applications (e.g., in educational testing), it can lead to misleading results in other settings.  For example, it is not uncommon to see a few members of the U.S.\ House of Representatives that are understood to be on opposite ends of the ideological spectrum vote together against a measure supported by the majority (e.g., see \citealp{Lewis2019a}, \citealp{Lewis2019b}, \citealp{yu2021spatial} and \citealp{DuckMayrMontgomery2022}).  In those circumstances, scaling methods that assume that response functions are monotonic will lead to estimates of the latent traits that make these extreme legislators appear centrists.

The literature on scaling models that can accommodate non-monotonic response functions is limited.  Probably, the best known example is the Generalized Graded Unfolding Model (GGUM), originally introduced in the psychology literature by \citet{RobertsDonoghueLaughlin2000}. GGUM ``folds'' the response function of a traditional graded response model, which allows individuals to exhibit lower response probabilities when an issue is further away from their preferences in either direction of the latent scale.  Computational methods for  GGUMs have been implemented in various \proglang{R} packages, including \pkg{GGUM} \citep{TendeiroCastroAlvarez2019}, which provides an interface to the original \proglang{Fortran} code developed for the \pkg{GGUM2004} software \citep{roberts2006ggum2004}, as well as \pkg{mirt} \citep{chalmers2012mirt} and \pkg{ScoreGGUM} \citep{KingRoberts2015}. Bayesian estimation of GGUMs using Markov chain Monte Carlo (MCMC) algorithms was first considered by \cite{JohnsonJunker2003b} and \cite{DeLaTorreStarkChernyshenko2006}. In particular, \cite{JohnsonJunker2003b} demonstrated that Bayesian estimates avoid the kind of bias to which frequentist estimates are prone when the model parameters lie at the end fo the latent scale.  MCMC algorithms for GGUMs have been implemented in packages \pkg{MCMC GGUM} \citep{WangDeLaTorreDrasgow2015} and \pkg{bmggum} \citep{TuZhangAngraveSun2021}.  However, the algorithms implemented in these packages can mix slowly and require substantial ad hoc tuning.  More recently, \citet{DuckMayrMontgomery2022} proposed a Metropolis-Coupled MCMC (MC3) method, available in the \pkg{bggum} package. This MC3 algorithm enhances the exploration of parameter spaces by allowing multiple parallel chains to exchange information while running at different temperatures.

Recently, \cite{lei2025novel} introduced probit unfolding models (PUMs) as an alternative to GGUMs. Unlike GGUMs, probit unfolding models can be directly motivated as spatial voting models, making their use especially appealing in political science applications. Furthermore, \cite{lei2025novel} provide substantial empirical evidence that probit unfolding models lead to complexity-adjusted fits that are at least as good as those of GGUMs in real applications.  This paper describes \pkg{pumBayes}, a new \proglang{R} package that enables Bayesian estimation of probit unfolding models using MCMC algorithm that require minimal or no tuning.  \pkg{pumBayes} can accommodate both static models in which the latent traits of the given individual is the same for all items, and dynamic models in which batches of observations are collected at different points in time and the latent is allowed to evolve slowly from one batch to another.  The package also includes various support functions that can be used to pre-process data, select hyperparameters, summarize output, and compute metrics of model fit.  While the package (and the illustrations) presented in this paper are geared towards political science applications, the methodology is quite general and can be applied to other settings (such as marketing or non-cognitive measurement applications) in which we might expect non-monotonic response functions.

The remainder of the paper is organized as follows:  Section \ref{sec:PUM} reviews probit unfolding models, including prior specification and the basics of posterior computation.  Section \ref{sec:usingPUM} provides two detailed illustrations of the use of \pkg{pumBayes}, one focused on the estimation of static probit unfolding models, and one focused on dynamic probit unfolding models.  The paper concludes with a brief discussion in Section \ref{sec:discussion}.

%% -- Manuscript ---------------------------------------------------------------

%% - In principle "as usual" again.
%% - When using equations (e.g., {equation}, {eqnarray}, {align}, etc.
%%   avoid empty lines before and after the equation (which would signal a new
%%   paragraph.
%% - When describing longer chunks of code that are _not_ meant for execution
%%   (e.g., a function synopsis or list of arguments), the environment {Code}
%%   is recommended. Alternatively, a plain {verbatim} can also be used.
%%   (For executed code see the next section.)

\section{The probit unfolding model}\label{sec:PUM}

The probit unfolding model (PUM for short, \citealp{lei2025novel}) is a one-dimensional factor analysis model closely related to item response models.  It can be motivated through the use of random utility functions.  To ground our presentation, we describe it using the usual political science  terminology.  In this context, we typically have $I$ legislators/judges who vote in favor or against $J$ issues.  PUM assumes that the (latent) preferences of legislator/judge $i$, denoted by $\beta_i$, belong to a one-dimensional Euclidean latent ``policy'' space.  It is common to refer to $\beta_i$ as the ``ideal point'' of legislator/judge $i$, as it represents their preferred policy.  Similarly, each issue has associated with it \textit{three} positions in the policy space, $\psi_{j,1}$, $\psi_{j,2}$ and $\psi_{j,3}$ such that either $\psi_{j,1} < \psi_{j,2} < \psi_{j,3}$ or $\psi_{j,3} < \psi_{j,2} < \psi_{j,1}$, where $\psi_{j,2}$ is associated with an affirmative vote, while both $\psi_{j,1}$ and $\psi_{j,3}$ are associated with a negative vote on issue $j$.  Legislators/judges choose among these three options on the basis of (random) utility functions of the form:
\begin{equation}\label{eq:utilities}
\begin{aligned}
U_{N-}(\beta_i, \psi_{j,1}) &= -(\beta_i - \psi_{j,1})^2 + \epsilon_{i,j,1}, \\
U_Y(\beta_i, \psi_{j,2}) &= -(\beta_i - \psi_{j,2})^2 + \epsilon_{i,j,2}, \\
U_{N+}(\beta_i, \psi_{j,3}) &= -(\beta_i - \psi_{j,3})^2 + \epsilon_{i,j,3},
\end{aligned}
\end{equation}
where $\epsilon_{i,j,t,1}$, $\epsilon_{i,j,t,2}$, and $\epsilon_{i,j,3}$ are independent standard Gaussian shocks.  Then, if we let $y_{i,j} \in \{0, 1\}$ represent the vote of legislator/judge $i$ on issue $j$, the probability of an affirmative vote is given by
\begin{align}
        P(y_{i,j} = 1 \mid \beta_i, & \psi_{j,1}, \psi_{j,2}, \psi_{j,3}) = P(\epsilon_{i,j,1} - \epsilon_{i,j,2} < \alpha_{j,1}(\beta_i - \delta_{j,1}), \epsilon_{i,j,3} - \epsilon_{i,j,2} < \alpha_{j,2}(\beta_i - \delta_{j,2})) \nonumber \\
        & = \int_{-\infty}^{\alpha_{j,1}(\beta_i-\delta_{j,1})} \int_{-\infty}^{\alpha_{j,2}(\beta_i-\delta_{j,2})} \frac{1}{2\sqrt{3\pi}} \exp\left\{-\frac{1}{3} \left( z_1^2 - z_1 z_2 + z_2^2 \right) \right\} dz_1 dz_2 \label{eq:theta_ij_def},
\end{align}
where $\alpha_{j,1} = 2(\psi_{j,2} - \psi_{j,1}), \alpha_{j,2} = 2(\psi_{j,2} - \psi_{j,3}),
\delta_{j,1} = (\psi_{j,1} + \psi_{j,2})/2$  and  $\delta_{j,2} = (\psi_{j,3} + \psi_{j,2})/2$.

Note that this formulation includes a 2-parameter probit IRT model as a limiting case.  Indeed, %if $\delta_{i,j} \to \infty$ then it is easy to see that
\begin{multline*}
        \lim_{\delta_{i,j} \to \infty} P(y_{i,j} = 1 \mid \beta_i, \psi_{j,1}, \psi_{j,2}, \psi_{j,3}) =
        \\  \int_{-\infty}^{\alpha_{j,2}(\beta_i-\delta_{j,2})} \frac{1}{2\sqrt{3\pi}} \exp\left\{-\frac{1}{2} z_1^2  \right\} dz_1 = \Phi\left( \alpha_{j,2}(\beta_i-\delta_{j,2}) \right) .
\end{multline*}

\subsection{Prior distributions}

\pkg{pumBayes} uses prior distributions for the issue-specific parameters $(\boldsymbol{\alpha}_j,\boldsymbol{\delta}_j)$ that are independent across $j=1, \ldots, J$ and take the form
\begin{multline*}
p(\bfalpha_j,\bfdelta_j) = \frac{1}{2}\text{TN}_2(\bfalpha_j\mid \bm{0}, \omega \mathbbm{I}_{2\times 2})\bm{1}(\alpha_{j,1}>0,\alpha_{j,2}<0)\text{N}_2(\bfdelta_j\mid \bfmu,\kappa\mathbbm{I}_{2\times 2})\\
+ \frac{1}{2}\text{TN}_2(\bfalpha_j\mid \bm{0}, \omega \mathbbm{I}_{2\times 2})\bm{1}(\alpha_{j,1}<0,\alpha_{j,2}>0)\text{N}_2(\bfdelta_j\mid -\bfmu,\kappa\mathbbm{I}_{2\times 2}) ,
\end{multline*}
where $\boldsymbol{\alpha}_j = (\alpha_{j,1}, \alpha_{j,2})$, $\boldsymbol{\delta}_j = (\delta_{j,1}, \delta_{j,2})$, $\text{N}_q( \cdot \mid \bm{a}, \bm{A})$ denotes the $q$-variate normal distribution with mean vector $\bm{a}$ and variance matrix $\bm{A}$ and, similarly, $\text{TN}_q( \cdot \mid \bm{a}, \bm{A})\bm{1}(\Omega)$ denotes the $q$-variate normal distribution with mean vector $\bm{a}$ and variance matrix $\bm{A}$ truncated to $\Omega$.

For the static version of the model, \pkg{pumBayes} uses a standard normal distribution as the prior for the ideal point, i.e., $\beta_i \sim N(0,1)$ independently for all $i=1, \ldots, I$.  For the dynamic version of the model, the ideal points are still assumed to be independent across individuals, but they are linked across time for a given individual using a first-order autoregressive process whose stationary distribution is again a standard normal distribution,
\begin{align*}
\beta_{i,t} &= \rho \beta_{i,t-1} + \nu_t,  & \nu_t &\sim \text{N} (0, 1-\rho^2), & 0 \le \rho < 1 .
\end{align*}
The autocorrelation parameter $\rho$ of this latent autoregressive process is then given a hyperprior (a Gaussian distribution truncated to the $(0,1)$ interval) and its value learned from the data.  This choice of prior distributions ensure that the parameters are identifiable to shifts and scalings of the latent policy space.  In order to ensure identifiability to reflections of the policy space, users must identify individual(s) in the data for whom the sign of their ideal point will be fixed to be positive.

\subsection{Posterior computation}

\pkg{pumBayes} relies on a MCMC algorithm to generate samples from the posterior distribution of the model.  The construction of the sampler involves two data augmentation tricks.  The first augmentation trick is reminiscent of that described in \cite{albert1993bayesian} and involves the introduction of vectors of latent variables, $\bfy_{i,j}^{*} = (y^{*}_{i, j, 1}, y^{*}_{i, j, 2}, y^{*}_{i, j, 3})$ for every $i=1\ldots,I$ and $j=1,\ldots,J$. The definition of these latent variables is tightly linked to the form of the utility functions in \eqref{eq:utilities}:
\begin{align*}
    y^{*}_{i, j, 1} & = - \alpha_{j,1}(\beta_{i} - \delta_{j,1}) + e_{i, j, 1}, \\
    y^{*}_{i, j, 2} & =  e_{i, j, 2}, \\
    y^{*}_{i, j, 3} & = - \alpha_{j,2}(\beta_{i} - \delta_{j,2}) + e_{i, j, 3},
\end{align*}
where $e_{i, j, 1}$, $e_{i, j, 2}$ and $e_{i, j, 3}$ are standard normal distributions. This augmentations ensures that most of the full conditional distributions belong to standard families from which direct sampling is possible.

The second augmentation trick breaks down the mixture prior on $\bfalpha_j$ and $\bfdelta_j$ into the two fully connected regions associated with their support.  In particular, for $j=1,\ldots,J$, we let $z_{j} = 1$ if and only if $\alpha_{j,1} > 0$ and $\alpha_{j,2} < 0$, and $z_{j} = -1$ otherwise.  Then, $\Pr(z_{i,j} = 1) = \Pr(z_{i,j} = -1) = 1/2$, and
\begin{align*}
    p(\bfalpha_j, \bfdelta_j \mid z_j)& \propto \begin{cases}
    \exp\left\{ -\frac{1}{2} \left(\frac{1}{\omega^2 }\bfalpha_j'\bfalpha_j +
    \frac{1}{\kappa^2}(\bfdelta_j - \bfvartheta)'(\bfdelta_j - \bfvartheta) \right)\right\} \mathbbm{1}(\alpha_{j,1}>0, \alpha_{j,2}<0) & z_j = 1 , \\
    \exp\left\{ -\frac{1}{2} \left(\frac{1}{\omega^2 }\bfalpha_j'\bfalpha_j +
    \frac{1}{\kappa^2}(\bfdelta_j + \bfvartheta)'(\bfdelta_j + \bfvartheta) \right)\right\} \mathbbm{1}(\alpha_{j,1}<0, \alpha_{j,2}>0) & z_j = -1. \\
    \end{cases}
\end{align*}

This second augmentation is important to ensure that the algorithm can fully explore the posterior distribution of $\bfalpha_j$ and, in particular, that it can move between the quadrant where $\alpha_{j,1} >0$ and $\alpha_{j,2}<0$, and that where $\alpha_{j,1} <0$ and $\alpha_{j,2}>0$.  To sample $z_j$, the algorithm relies on two Metropolis-Hasting steps with different relative frequencies.  For the first of these moves, proposed values are generated by flipping the signs of $\bfalpha_j$ and $\bfdelta_j$, while for the second move the proposed values for $\bfalpha_j$ and $\bfdelta_j$ are generated from the prior conditioned on the proposed quadrant.  The first of these moves is expected to be most effective for items for which $\bfalpha_j$ is close to $\mathbf{0}$ (e.g., quasi-unanimous votes in our political science application).  The number of $(\bfalpha_j, \bfdelta_j)$ pairs in this category is typically relatively low, but when the move is useful, the acceptance probability tends to be very high.  Hence, it is often appropriate to attempt it relatively infrequently.  On the other hand, the second move is most useful when the posterior distribution of $(\alpha_j, \delta_j)$ is clearly bimodal.  This move tends to be useful for a larger number of $(\bfalpha_j, \bfdelta_j)$ pairs, but the proposals also tend to have more moderate acceptance probabilities.  Hence, we propose to perform it relatively more frequently. Full details of the algorithm can be found in \cite{lei2025logit} and \cite{lei2025novel}.

\section{Using pumBayes}\label{sec:usingPUM}

\subsection{Installing the package}

\pkg{pumBayes} is available from \url{https://www.github.com/SkylarShiHub/pumBayes} and can be installed using the \code{install\_github} function in the \pkg{devtools} package:

\begin{Schunk}
\begin{Sinput}
R> if (!requireNamespace("pumBayes", quietly = TRUE)) {
+    devtools::install_github("SkylarShiHub/pumBayes")
+  }
R> library(pumBayes)
\end{Sinput}
\end{Schunk}

Since part of the package is written in the \proglang{C++} language, \pkg{pumBayes} relies on the external package \pkg{Rcpp} for its core functionality. Its two main functions are \code{sample\_pum\_static} (which, as the name suggests, generates posterior samples from the static version of PUM) and \code{sample\_pum\_dynamic} (which generates samples from the dynamic version).%  The package also includes various support functions that can be used to pre-process data, explore the impact of hyperparameters, summarize output, and compute metrics of model fit.

\subsection{Fitting static probit unfolding models using pumBayes}

We start by demonstrating the use of the function \code{sample\_pum\_static} to fit the static version of PUM.  \code{sample\_pum\_static} can handle data in the form of either a \code{rollcall} object from the \pkg{pscl} package \citep{jackman2024pscl}, or in the form of a logical matrix where \code{TRUE} values correspond to affirmative votes, \code{FALSE} values correspond to negative votes, and \code{NA}s  correspond to any type of missing data.  If a matrix object is used as input, the names of the rows are assumed to correspond to the legislators names. % If a \code{rollcall} object is used as input, the votes for ``Yea'' will be encoded as 1 and for ``Nay'' will be encoded as 0. Other situations encoded as will be ``NA''.

For our demonstration, we use \pkg{pumBayes} to recover legislators' preferences from roll-call voting data from the 116\textsuperscript{th} U.S.\ House of Representatives.  The data can be downloaded from \url{https://voteview.com} using the \pkg{pscl} package (see below), and it is also included in \pkg{pumBayes} as \code{h116}.

\begin{Schunk}
\begin{Sinput}
R> require(pscl)
R> h116 = readKH("https://voteview.com/static/data/out/votes/H116_votes.ord",
+                desc = "116th U.S. House of Representatives")
\end{Sinput}
\begin{Soutput}
Attempting to read file in Keith Poole/Howard Rosenthal (KH) format.
Attempting to create roll call object
116th U.S. House of Representatives 
452 legislators and 952 roll calls
Frequency counts for vote types:
rollCallMatrix
     0      1      3      6      7      9 
 18195 258570      1 138011     84  15443 
\end{Soutput}
\end{Schunk}

The object \code{h116} is of class \code{rollcall} (which is defined in \pkg{pscl}).  For our purposes, the key component of any \code{rollcall} object is \code{votes} which, as the name suggests, contains the outcomes of the votes cast by each legislator who served in the House during this period.  These outcomes are typically encoded as 1, 2 or 3 for an affirmative vote (``Yea'', ``Paired Yea'' and ``Announced Yea''), 4, 5, or 6 for a negative vote (``Announced Nay'', ``Paired Nay'' and ``Nay''), and 0, 7, 8, 9 or \code{NA} for different types of missing values (``Not Yet a Member'', ``Present'' and ``Not Voting'').  The component \code{codes} of the \code{rollcall} object typically includes an explanation of these codes.

\begin{Schunk}
\begin{Sinput}
R> print(h116$votes[1: 6, 1: 6])
\end{Sinput}
\begin{Soutput}
                  Vote 1 Vote 2 Vote 3 Vote 4 Vote 5 Vote 6
TRUMP (R NA)           0      0      0      0      0      0
BYRNE (R AL-1)         6      6      6      1      6      6
ROBY (R AL-2)          6      6      6      1      6      6
ROGERS (R AL-3)        6      6      6      1      6      6
ADERHOLT (R AL-4)      6      6      6      1      6      6
BROOKS (R AL-5)        6      6      6      1      6      6
\end{Soutput}
\begin{Sinput}
R> print(h116$codes)
\end{Sinput}
\begin{Soutput}
$yea
[1] 1 2 3

$nay
[1] 4 5 6

$notInLegis
[1] 0

$missing
[1] 7 8 9
\end{Soutput}
\end{Schunk}

Another useful component of a \code{rollcall} object is \code{legis.data}, which contains information about the legislators involved with this dataset.

\begin{Schunk}
\begin{Sinput}
R> print(h116$legis.data[1: 6, ])
\end{Sinput}
\begin{Soutput}
                  state icpsrState cd icpsrLegis party partyCode
TRUMP (R NA)       <NA>         99  0      99912     R       200
BYRNE (R AL-1)       AL         41  1      21376     R       200
ROBY (R AL-2)        AL         41  2      21192     R       200
ROGERS (R AL-3)      AL         41  3      20301     R       200
ADERHOLT (R AL-4)    AL         41  4      29701     R       200
BROOKS (R AL-5)      AL         41  5      21193     R       200
\end{Soutput}
\end{Schunk}

Typically, datasets downloaded from \url{https://voteview.com} need to be preprocessed before they can be analyzed.  The function \code{preprocess\_rollcall} in \pkg{pumBayes} can handle transformations of the original \code{rollcall} object, including the removal of pre-specified legislators or votes, or the merging of pairs of legislators. In the specific case of \code{h116}, we will use this function to perform the following operations on the data:
\begin{itemize}
\item The President of the United States at the time (Donald Trump) needs to be removed.  The following code identifies the row index in the matrix of votes that corresponds to him (note that there are no other legislators in this House that share the last name TRUMP):

\begin{Schunk}
\begin{Sinput}
R> legis.to.remove = grep("TRUMP", rownames(h116$votes))
\end{Sinput}
\end{Schunk}

\item In the analysis of congressional voting records, lopsided (unanimous or quasi-unanimous) votes are often removed before scaling because they provide little or no information about legislators preferences. In this cases, we are interested in removing only completely unanimous votes.

\begin{Schunk}
\begin{Sinput}
R> lop.issue = 0
\end{Sinput}
\end{Schunk}

\item Two legislators (Justin Amash, representing MI-3, and Paul Mitchell, representing MI-10) left the Republican party during this period to become independents. Hence, each of them appears twice in \code{h116}. For the purpose of this analysis, we merge their records before and after becoming independents and label the combined voting records for the whole 116\textsuperscript{th} House as such.

\begin{Schunk}
\begin{Sinput}
R> legis.to.combine = list(grep("AMASH", rownames(h116$votes)),
+                          grep("MITCHELL", rownames(h116$votes)))
R> legis.to.combine.party = c("I", "I")
\end{Sinput}
\end{Schunk}

\item Finally, after all the above filters are applied, we are interested in removing any legislator who missed more than 60\% of the remaining votes from the dataset.

\begin{Schunk}
\begin{Sinput}
R> lop.leg = 0.6
\end{Sinput}
\end{Schunk}
\end{itemize}

All of the above parameters are bound into a list that is passed as an argument to the function \code{preprocess\_rollcall}.  This function returns a ``clean'' version of the original \code{rollcall} object:

\begin{Schunk}
\begin{Sinput}
R> control = list(leg_rm = legis.to.remove,
+                 combine_leg_index = legis.to.combine,
+                 combine_leg_party = legis.to.combine.party,
+                 lop_leg = lop.leg, lop_issue = lop.issue)
R> h116.c = preprocess_rollcall(h116, control)
\end{Sinput}
\end{Schunk}

The final step before being able to fit the model is to select the hyperparameters. The default values used by \pkg{pumBayes} correspond to $\bfmu = (-2, 10)'$, $\omega = 5$ and $\kappa = \sqrt{10}$.  A way to evaluate the suitability of these default values is to visualize the implied prior on the probability of an affirmative outcome obtained from Equation \eqref{eq:theta_ij_def}.  The function \code{tune\_hyper} allows us to generate a random sample from this implied distribution, and a histogram of these samples provides a convenient visualization.
\begin{Schunk}
\begin{Sinput}
R> hyperparams = list(beta_mean = 0, beta_var = 1,
+                     alpha_mean = c(0, 0), alpha_scale = 5,
+                     delta_mean = c(-2, 10), delta_scale = sqrt(10))
R> theta = tune_hyper(hyperparams, n_leg = 1000, n_issue = 1000)
R> hist(theta, freq = FALSE, xlab = expression(theta[list(i, j)]),
+       main = "Default prior")
\end{Sinput}
\end{Schunk}
\begin{figure}[t!]
\centering
\includegraphics[width=0.65\textwidth]{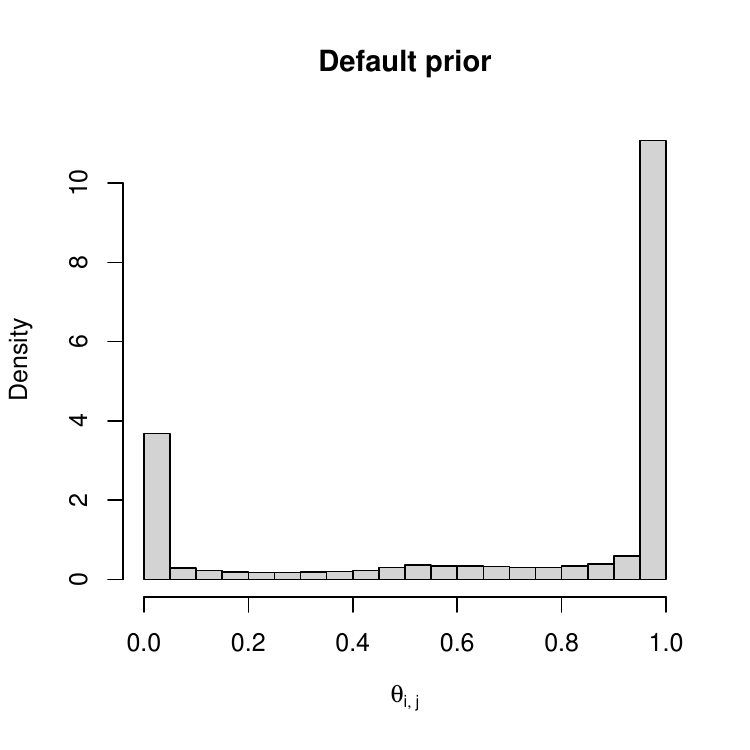}
\caption{\label{fig:tune_hyper1} Default prior distribution for affirmative voting probability.}
\end{figure}

As Figure \ref{fig:tune_hyper1} shows, the default hyperparameter values imply a prior that places most of its mass close to either 0 and 1.  This in turn implies an assumption that, for most votes, the majority of legislators are quite certain about how they will vote.  This seems most appropriate in the context of this application. The prior also places slightly more mass close to 1 than it does close to 0. Again, this is appropriate, as the agenda-setting powers of the majority means that most motions put to a vote in the U.S. House tend to pass, and therefore, receive more affirmative than negative votes.

\begin{figure}[t!]
\centering
\includegraphics[width=0.65\textwidth]{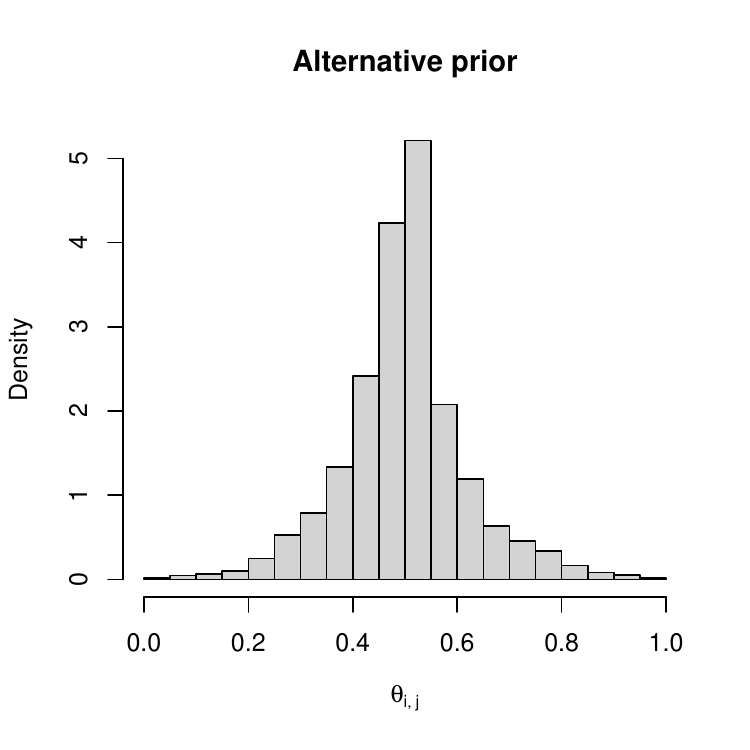}
\caption{\label{fig:tune_hyper2} Alternative prior distribution for affirmative voting probability.}
\end{figure}

The function \code{tune\_hyper} can also be used to explore the implications of alternative prior specifications. For example, a prior that is concentrated around 0.5 can be obtained by setting $\bfmu = (-40, 0)'$, $\omega = 0.1$ and $\kappa = 5$ (please see Figure \ref{fig:tune_hyper2}).
\begin{Schunk}
\begin{Sinput}
R> hyperparams.alt = list(beta_mean = 0, beta_var = 1,
+                         alpha_mean = c(0, 0), alpha_scale = 0.1,
+                         delta_mean = c(-40, 0), delta_scale = 5)
R> theta.alt = tune_hyper(hyperparams.alt, n_leg = 1000, n_issue = 1000)
R> hist(theta.alt, freq = FALSE, xlab = expression(theta[list(i, j)]),
+       main = "Alternative prior")
\end{Sinput}
\end{Schunk}

We are now ready to invoke the function \code{sample\_pum\_static} to sample from the posterior distribution.  We aim at generating a total of 400,000 samples, of which the first 200,000 will be discarded as part of the burn-in period. These samples are obtained by thinning a longer chain every 10 iterations.
\begin{Schunk}
\begin{Sinput}
R> control = list(num_iter = 400000, burn_in = 200000, keep_iter = 10,
+                 flip_rate = 0.1)
R> h116.c.pum = sample_pum_static(h116.c, hyperparams, control,
+                                 pos_leg = grep("SCALISE",
+                                              rownames(h116.c$votes)),
+                                 verbose = FALSE)
\end{Sinput}
\end{Schunk}

The ideological ranks implied by the ideal points are the key quantities of interest. These can be obtained using the function \code{post\_rank}.

\begin{Schunk}
\begin{Sinput}
R> h116.c.beta.pum.rank = post_rank(h116.c.pum$beta, c(0.05, 0.5, 0.95))
R> head(h116.c.beta.pum.rank)
\end{Sinput}
\begin{Soutput}
               name 0.05 0.5 0.95
1    BYRNE (R AL-1)  395 405  413
2     ROBY (R AL-2)  271 280  293
3   ROGERS (R AL-3)  334 346  363
4 ADERHOLT (R AL-4)  305 318  330
5   BROOKS (R AL-5)  425 428  429
6   PALMER (R AL-6)  388 398  407
\end{Soutput}
\end{Schunk}

Note that the posterior median rank for Mo Brooks, (who represented Alabama's 5\textsuperscript{th} district) is 428, placing him among the most conservative Republicans in the House during this period.  On the other hand, the posterior median rank for Martha Roby (representing Alabama's 2\textsuperscript{nd} district) is 280, placing her as a mainstream, even centrist Republican.

It is illustrative to compare the ideological ranks estimated by PUM with those estimated using traditional approaches such as IDEAL \citep{JackmanMultidimensionalAnalysisRoll2001}, which is implemented in the \pkg{pscl} package:

\begin{Schunk}
\begin{Sinput}
R> cl = constrain.legis(h116.c, x = list("CLYBURN" = -1, "SCALISE" = 1),
+                       d = 1)
R> h116.c.ideal = ideal(h116.c, d = 1, priors = cl, startvals = cl,
+                       maxiter = 20000, thin = 50, burnin = 10000,
+                       store.item = TRUE)
\end{Sinput}
\end{Schunk}

An effective way to visualize this comparison is a scatterplot of the posterior median ranks estimated by the two models.

\begin{Schunk}
\begin{Sinput}
R> h116.c.beta.ideal.rank = post_rank(h116.c.ideal$x[, , 1],
+                                     c(0.05, 0.5, 0.95))
R> partycolors = ifelse(h116.c$legis.data$party == "R", "red",
+                       ifelse(h116.c$legis.data$party == "D",
+                              "blue", "green"))
R> partysymbols = ifelse(h116.c$legis.data$party == "R", 17,
+                        ifelse(h116.c$legis.data$party == "D", 16, 8))
R> leg.to.plot = c("OCASIO-CORT (D NY-14)", "TLAIB (D MI-13)",
+                  "PRESSLEY (D MA-7)", "OMAR (D MN-5)")
R> ind.leg.to.plot = match(leg.to.plot, rownames(h116.c$votes))
R> plot(h116.c.beta.ideal.rank$`0.5`, h116.c.beta.pum.rank$`0.5`,
+       xlab = "Median Rank order (IDEAL)", ylab = "Median Rank Order (PUM)",
+       col = partycolors, pch = partysymbols, cex.lab = 1)
R> text(h116.c.beta.ideal.rank$`0.5`[ind.leg.to.plot],
+       h116.c.beta.pum.rank$`0.5`[ind.leg.to.plot] + 5,
+       labels = leg.to.plot, pos = c(4, 1, 3, 2), cex = 0.8)
\end{Sinput}
\end{Schunk}

\begin{figure}[t!]
\centering
\includegraphics[width=0.65\textwidth]{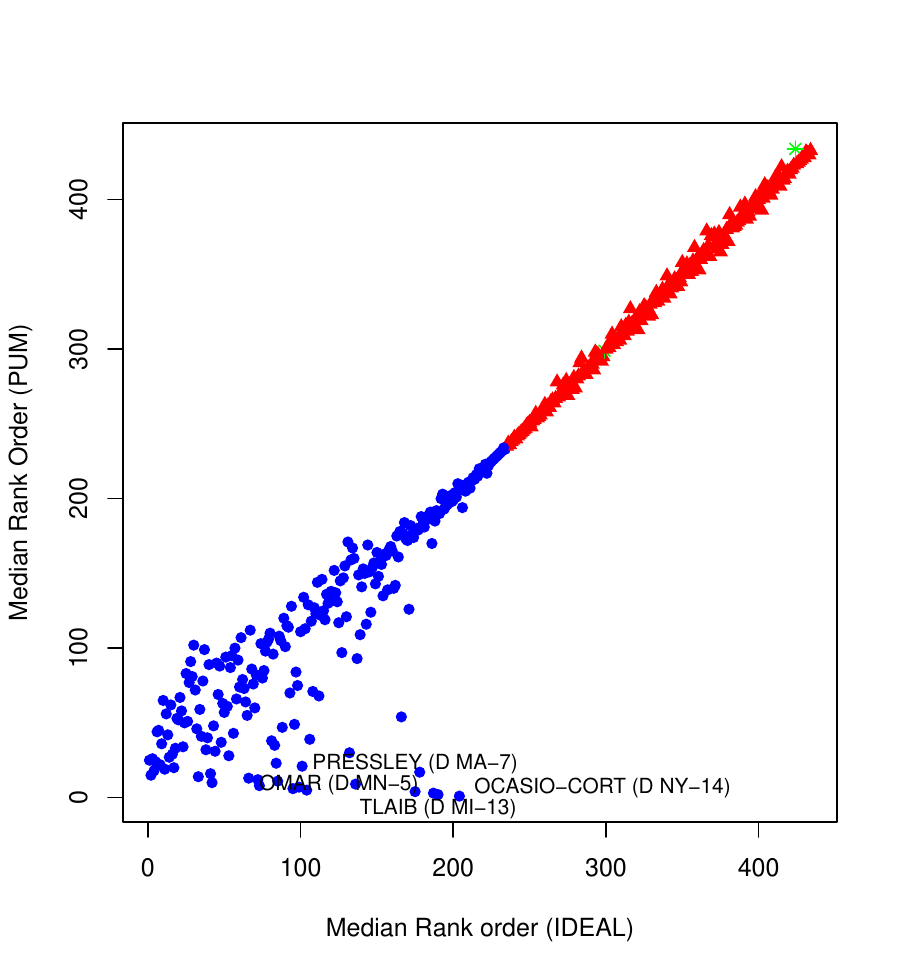}
\caption{\label{fig:rank_plot1} Comparison of median ranks: PUM vs. IDEAL.}
\end{figure}

As Figure \ref{fig:rank_plot1} shows, the rankings generated by PUM can differ significantly from those generated by IDEAL, especially for Democratic legislators. The top four discrepancies are for the members of the ``Squad'': Representatives Alexandria Ocasio-Cortez (NY), Ilhan Omar (MN), Ayanna Pressley (MA), and Rashida Tlaib (MI). IDEAL positions them as centrists, while PUM places them among the most liberal in the Democratic caucus. Given the Squad's advocacy for progressive policies, PUM's rankings seem much more accurate (e.g., see \citealp{Lewis2019b,Lewis2019a}).

\begin{figure}[t!]
\centering
\includegraphics[width=0.65\textwidth]{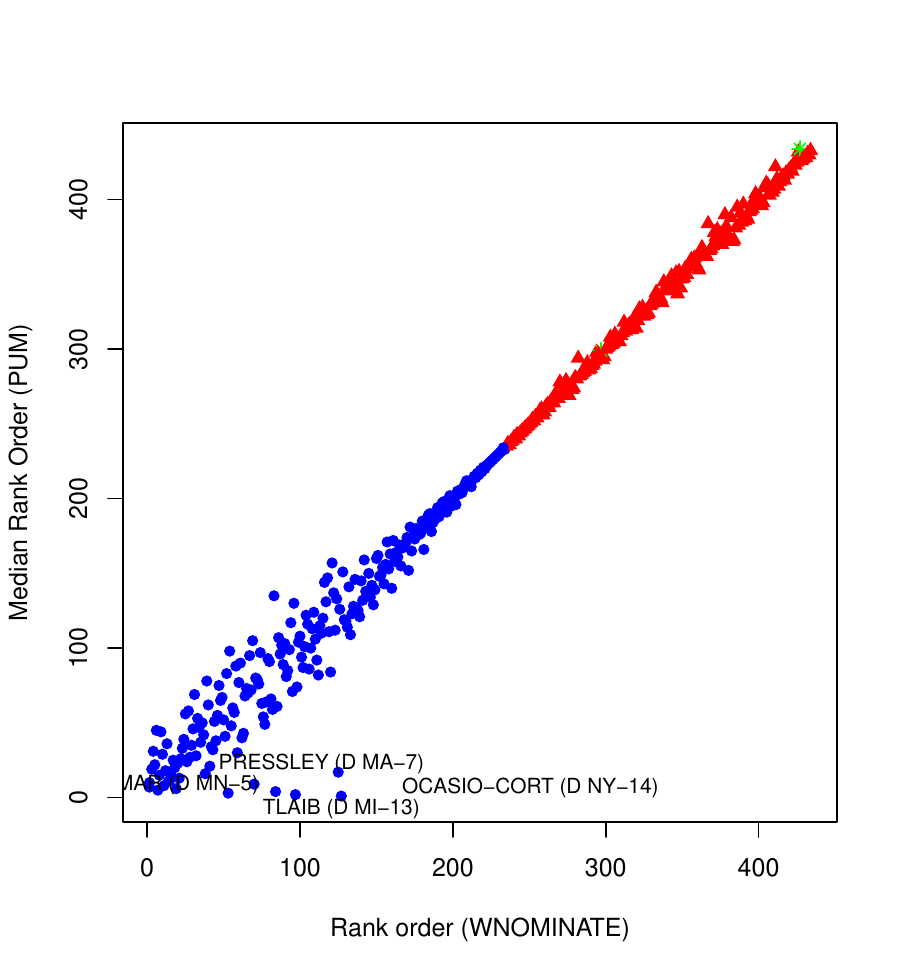}
\caption{\label{fig:rank_plot2} Comparison of median ranks: PUM vs. WNOMINATE.}
\end{figure}

The results look similar if we instead compare against the ranks generated using the \pkg{wnominate} package (see Figure \ref{fig:rank_plot2}):

\begin{Schunk}
\begin{Sinput}
R> library(wnominate)
R> h116.c.wnominate = wnominate(h116.c, dims = 1,
+                          polarity = grep("SCALISE",
+                                          rownames(h116.c$legis.data)))
R> beta.wnom.rank = rank(h116.c.wnominate[["legislators"]][["coord1D"]])
\end{Sinput}
\end{Schunk}
\begin{Schunk}
\begin{Sinput}
R> plot(beta.wnom.rank, h116.c.beta.pum.rank$`0.5`,
+       xlab = "Rank order (WNOMINATE)", ylab = "Median Rank Order (PUM)",
+       col = partycolors, pch = partysymbols, cex.lab = 1)
R> text(beta.wnom.rank[ind.leg.to.plot] + 30,
+       h116.c.beta.pum.rank$`0.5`[ind.leg.to.plot] + 5,
+       labels = leg.to.plot, pos = c(4, 1, 3, 2), cex = 0.8)
\end{Sinput}
\end{Schunk}

As we mentioned in the introduction, the differences in the estimates generated by these models are due to the fact that PUM allows for both monotonic and non-monotonic response functions.  To illustrate this, we construct estimates of the response functions associated with three separate votes using the function \code{item\_char} in \pkg{pumBayes}:

\begin{Schunk}
\begin{Sinput}
R> response.curve.pum1 = item_char(vote_num = 5, x = c(-4, 2), h116.c.pum)
R> response.curve.pum2 = item_char(vote_num = 9, x = c(-4, 2), h116.c.pum)
R> response.curve.pum3 = item_char(vote_num = 6, x = c(-4, 2), h116.c.pum)
R> item.list = list(response.curve.pum1, response.curve.pum2,
+                   response.curve.pum3)
R> vote.num.list = c(5, 9, 6)
R> par(mfrow = c(1, 3), mai = c(1, 1.1, 1, 1), bty = "n")
R> for (i in seq_along(item.list)) {
+    data = item.list[[i]]
+    plot(NA, xlim = range(data$beta_samples),
+         ylim = range(data$ci_lower, data$ci_upper),
+         xlab = expression(beta[i]), ylab = "Probability of voting 'Yes'",
+         main = paste0("Vote ", vote.num.list[i]), xaxt = "n",
+         cex.axis = 1.8, cex.main = 3, cex.lab = 2.3)
+    axis(1, at = c(-4, -2, 0, 2), labels = c(-4, -2, 0, 2), cex.axis = 2)
+    polygon(c(data$beta_samples, rev(data$beta_samples)),
+            c(data$ci_upper, rev(data$ci_lower)), col = "grey80", border=NA)
+    lines(data$beta_samples, data$means, lwd = 1.5, col = "black")
+  }
R> par(mfrow = c(1, 1))
\end{Sinput}
\end{Schunk}

\begin{figure}[t!]
\centering
\includegraphics[width=1\textwidth]{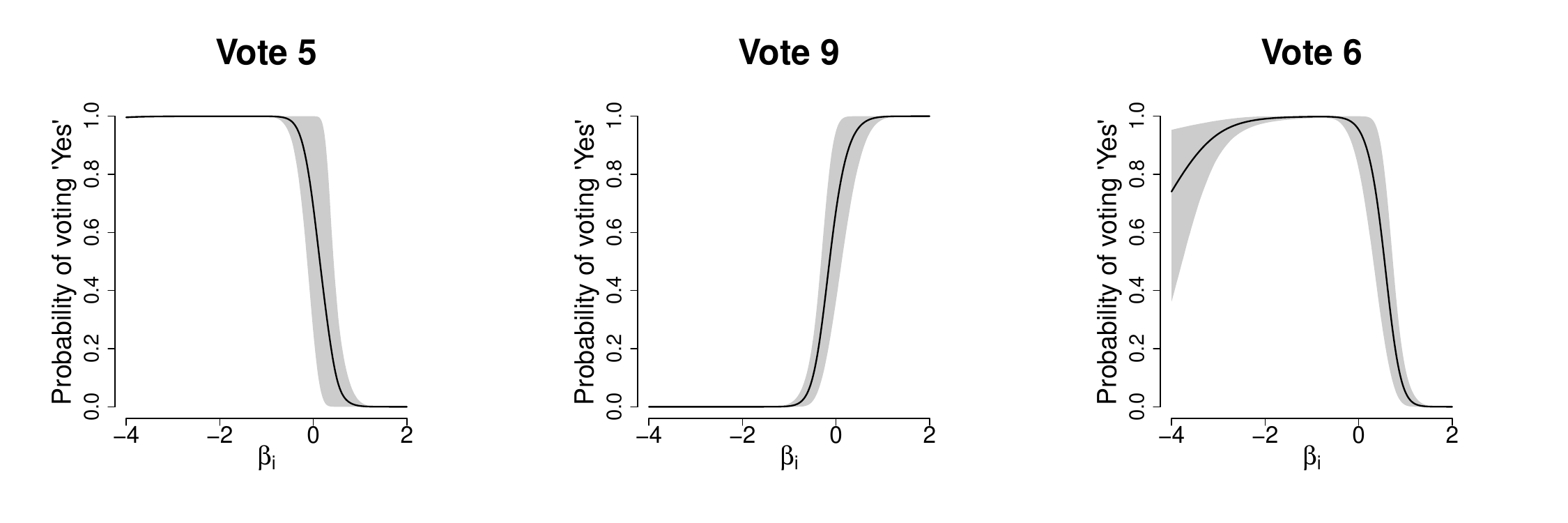}
\caption{\label{fig:item_response_curve} Estimated response functions for selected votes.}
\end{figure}

As Figure \ref{fig:item_response_curve} shows, PUM can capture both monotonic and non-monotonic response functions depending on the information contained in the observed data. Vote 5 (clerk session vote number 6) and vote 9 (clerk session vote number 10) are examples of votes with monotonic response functions, while vote 6 (clerk session vote number 7) is an example of a vote with a non-monotonic response function.

Comparisons between PUM and other Bayesian models can be carried out using a blocked version of the Watanabe-Akaike Information Criterion (WAIC, \citealp{watanabe2010asymptotic, watanabe2013widely, gelman2014understanding}).  For a given House and model, the  WAIC is given by
\begin{multline}\label{eq:lWAIC}
    WAIC = -2\left[\sum_{i=1}^{I} \log\left( \textrm{E}_{\textrm{post}} \left\{
    \prod_{j = 1}^J \theta_{i,j}^{y_{i,j}} \left[1 -\theta_{i,j}\right]^{1 - y_{i,j}} \right\} \right) \right. \\
     \left. - \sum_{i=1}^{I}  \textrm{var}_{\textrm{post}}\left\{
    \sum_{j = 1}^{J} \left[ y_{i,j} \log \theta_{i,j} + (1-y_{i,j}) \log(1-\theta_{i,j}) \right] \right\} \right] ,
\end{multline}
where $\theta_{i, j}$ represents the probability that legislator $i$ votes ``Aye'' in issue $j$. Lower values of WAIC provide evidence in favor of that particular model. Like the Akaike Information Criterion and Bayesian Information Criterion, the WAIC attempts to balance model fit with model complexity. However, unlike these two criteria, WAIC is appropriate for hierarchical setting and is invariant to reparametrizations of the model.

To calculate WAIC values, we first calculate the probability array of voting ``Aye'' using functions \code{predict\_pum} and \code{predict\_ideal} contained in \pkg{pumBayes}. The output of each of these functions in a three-dimensional array, where the dimensions correspond to members, issues and MCMC iterations. Then the function \code{calc\_waic} can be used to evaluate the complexity-adjusted fit of both PUM and IDEAL models using Equation \eqref{eq:lWAIC}:
\begin{Schunk}
\begin{Sinput}
R> h116.c.pum.predprob = predict_pum(h116.c, post_samples = h116.c.pum)
R> h116.c.ideal.predprob = predict_ideal(h116.c, h116.c.ideal)
R> h116.c.pum.waic = calc_waic(h116.c, prob_array = h116.c.pum.predprob)
R> h116.c.ideal.waic = calc_waic(h116.c, prob_array = h116.c.ideal.predprob)
\end{Sinput}
\end{Schunk}
\begin{Schunk}
\begin{Sinput}
R> print(h116.c.pum.waic)
\end{Sinput}
\begin{Soutput}
[1] 79351.92
\end{Soutput}
\begin{Sinput}
R> print(h116.c.ideal.waic)
\end{Sinput}
\begin{Soutput}
[1] 83649.23
\end{Soutput}
\end{Schunk}

These WAIC scores clearly indicate that PUM vastly outperforms IDEAL on this particular dataset.

\subsection{Fitting dynamic probit unfolding models using pumBayes}

We now demonstrate the use of the function \code{sample\_pum\_dynamic} for estimating time-varying preferences. Data is passed into \code{sample\_pum\_dynamic} through two objects.  The first is a logical matrix containing the vote outcomes and in which \code{TRUE} values correspond to affirmative votes, \code{FALSE} values to negative votes, and \code{NA}s correspond to any type of missing data.  This is the same as one of the formats accepted by \code{sample\_pum\_static}.  The second object is a vector that indicates the time period to which each vote belongs.

For our illustration, we focus on voting data for the U.S.\ Supreme Court between 1937 and 2021, which is available for download at \url{http://mqscores.wustl.edu/replication.php} and is included in our package as part of the object \code{scotus.1937.2021}. This dataset includes the voting outcomes on non-unanimous decision, with the outcome being encoded as 1 (\code{TRUE}) if the judge voted to reverse the lower court decision, and as 0 (\code{FALSE}) otherwise.

\begin{Schunk}
\begin{Sinput}
R> data(scotus.1937.2021)
R> mqVotes = array(as.logical(1 - mqVotes), dim = dim(mqVotes),
+                  dimnames = list(rownames(mqVotes), colnames(mqVotes)))
R> print(mqVotes[1: 6, 1: 8])
\end{Sinput}
\begin{Soutput}
                 3     4     6    10    12    13    15    22
CEHughes2    FALSE FALSE FALSE FALSE FALSE FALSE FALSE FALSE
JCMcReynolds  TRUE  TRUE  TRUE  TRUE FALSE FALSE FALSE  TRUE
LDBrandeis   FALSE FALSE FALSE FALSE FALSE FALSE  TRUE FALSE
GSutherland   TRUE  TRUE FALSE  TRUE FALSE FALSE FALSE  TRUE
PButler       TRUE  TRUE FALSE FALSE FALSE FALSE FALSE  TRUE
HFStone      FALSE FALSE FALSE FALSE  TRUE  TRUE  TRUE FALSE
\end{Soutput}
\end{Schunk}

The syntax for \code{sample\_pum\_dynamic} is very similar to that for \code{sample\_pum\_static}.  Furtheremore, note that we use the same hyperparameters for $(\bfalpha_j , \bfdelta_j)$ as we did in the static case:
\begin{Schunk}
\begin{Sinput}
R> hyperparams = list(alpha_mean = c(0, 0), alpha_scale = 5,
+                     delta_mean = c(-2, 10), delta_scale = sqrt(10),
+                     rho_mean = 0.9, rho_sigma = 0.04)
R> control = list(num_iter = 600000, burn_in = 200000, keep_iter = 20,
+                 flip_rate = 0.1, sd_prop_rho = 0.1)
R> sign.refs = list(pos_inds = c(39, 5), neg_inds = c(12, 29),
+                   pos_year_inds = list(1: 31, 1),
+                   neg_year_inds = list(1: 29, 1: 24))
R> scotus.pum = sample_pum_dynamic(mqVotes, mqTime, hyperparams, 
+                                  control, sign.refs, verbose = FALSE)
\end{Sinput}
\end{Schunk}

The path of the ideal points is often a quantity of interest.  For example, we might want to plot their posterior mean and associated $95\%$ credible intervals for a few selected Justices.
\begin{Schunk}
\begin{Sinput}
R> scotus.beta = scotus.pum$beta
R> names = c("Hugo Black", "Antonin Scalia", "Warren E. Burger")
R> terms = list(seq(1937, 1970), seq(1986, 2015), seq(1969, 1985))
R> labels = c("HLBlack", "AScalia", "WEBurger")
R> colors = c("blue", "red", "black")
R> polygon_colors = c("slategray1", "lightpink", "gray")
R> par(mai = c(1, 1.2, 1, 1))
R> plot(1937: 2015, rep(NA, length(1937: 2015)), type = "n", 
+       xlim = c(1937, 2015), ylim = c(-8, 5), xlab = "Year", 
+       ylab = expression(beta[list(i, t)]), cex.lab = 1.5, cex.axis = 1.5)
R> for (i in seq_along(terms)) {
+    ind = grep(labels[i], colnames(scotus.beta))
+    postmeans = colMeans(scotus.beta[, ind])
+    quantiles = apply(scotus.beta[, ind], 2, quantile, c(0.05, 0.975))
+    polygon(c(terms[[i]], rev(terms[[i]])),
+            c(quantiles[1, ], rev(quantiles[2, ])),
+            col = polygon_colors[i], border = "NA")
+    lines(terms[[i]], postmeans, lwd = 2, col = colors[i])
+  }
R> legend(1980, -2, names, lwd = 2, col = colors, bty = "n", cex = 1.5)
\end{Sinput}
\end{Schunk}

\begin{figure}[t!]
\centering
\includegraphics[width=0.9\textwidth]{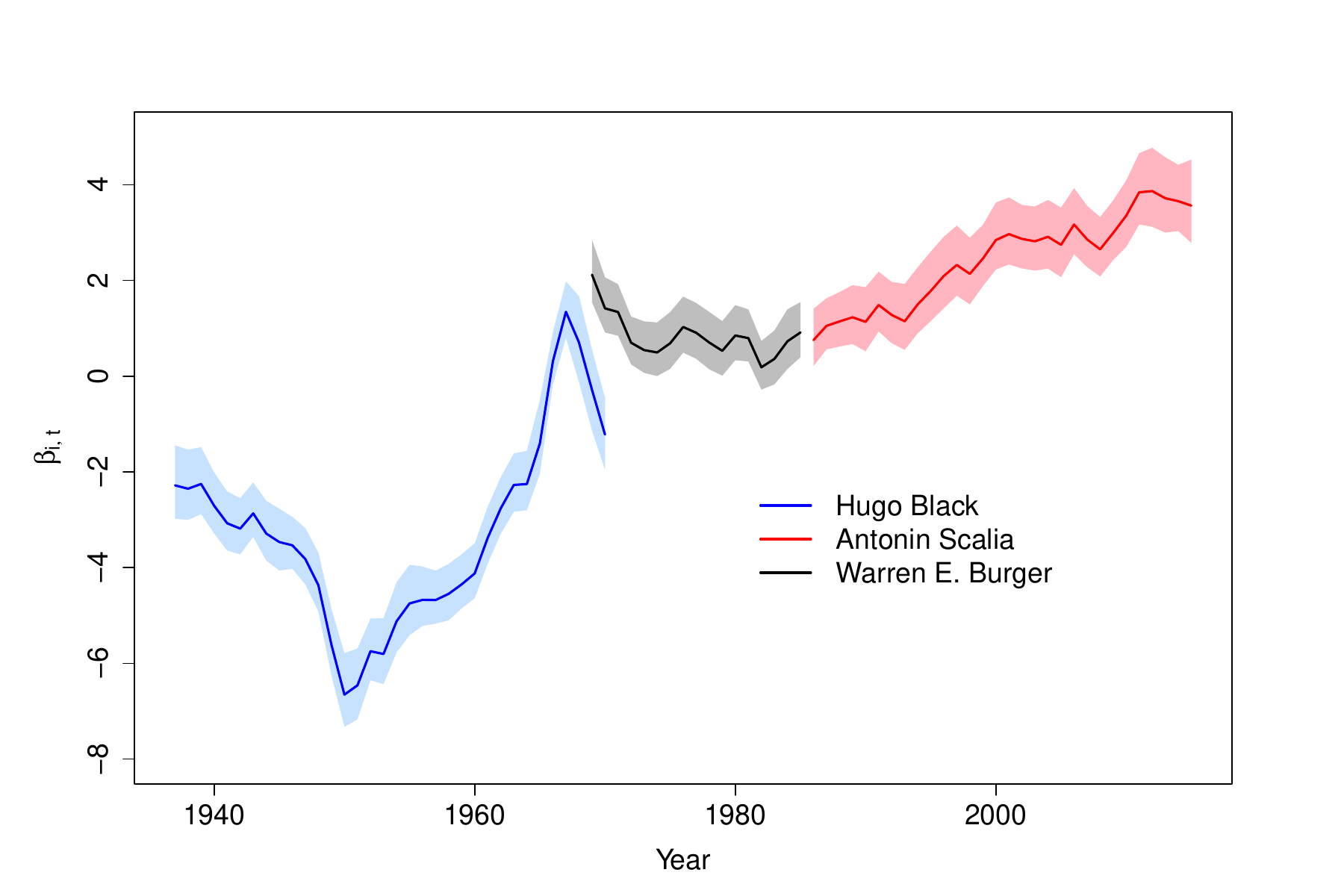}
\caption{\label{fig:ideal_points_paths} Evolution of justice preferences over time.}
\end{figure}

Figure \ref{fig:ideal_points_paths} suggests that Antonin Scalia's preferences steadily become more positive (``conservative'') over his time in the Court.  Hugo Black's preferences also seem to have evolved substantially over time but, in his case, his first became more negative  (`liberal'') and then more positive (``conservative'').  In contrast to these two Justices, Warren E. Burger's preferences remained relatively steady during his term as Chief Justice.

\begin{figure}[t!]
\centering
\includegraphics[width=1\textwidth]{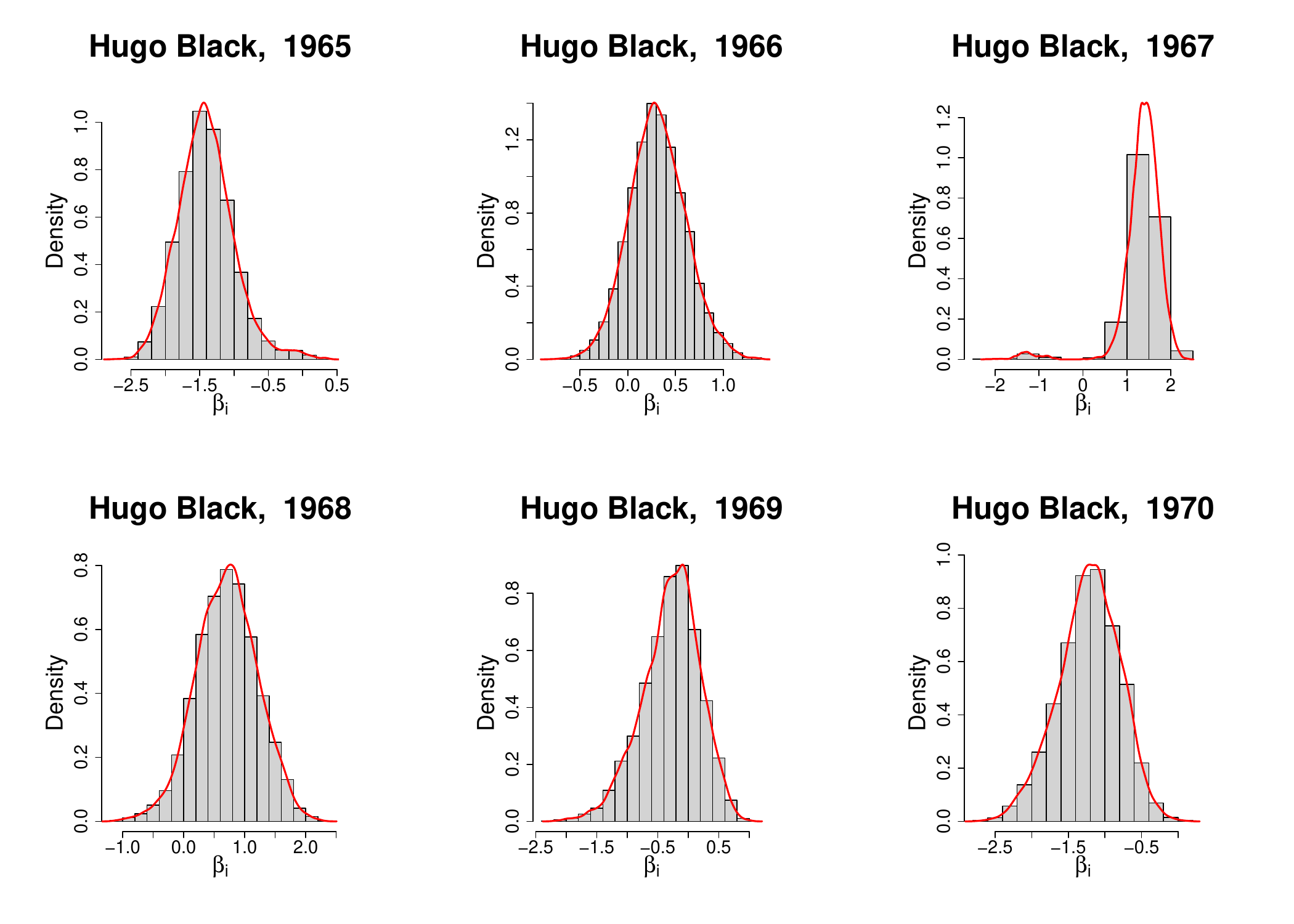}
\caption{\label{fig:ideal_points_Black_hist} Posterior distributions for justice Hugo Black's preferences.}
\end{figure}

A salient feature of PUM is that it can sometimes produce posterior distributions for the ideal points that are bimodal. Hugo Black's preferences during 1967 are an example  (see Figure \ref{fig:ideal_points_Black_hist}).  We note that this bimodality is consistent with results in reported in \cite{lei2024dynamic} using an alternative dynamic model based on a circular policy space and is not the result of lack of convergence in the MCMC algorithm.
\begin{Schunk}
\begin{Sinput}
R> scotus.beta.Black = scotus.beta[,grep("HLBlack",colnames(scotus.beta))]
R> term_Black = seq(1937,1970)
R> par(mfrow = c(2, 3), mai = c(1, 1.1, 1, 1))
R> for(i in 29: 34){
+    hist(scotus.beta.Black[, i], probability = TRUE,
+         main = paste("Hugo Black, ", term_Black[i]),
+         ylim = c(0, max(density(scotus.beta.Black[, i])$y)),
+         xlab = expression(beta[i]), cex.axis = 1.8, 
+         cex.main = 3, cex.lab = 2.3)
+    lines(density(scotus.beta.Black[, i]), col = "red", lwd = 2)
+  }
R> par(mfrow = c(1, 1))
\end{Sinput}
\end{Schunk}

\begin{figure}[t!]
\centering
\includegraphics[width=0.65\textwidth]{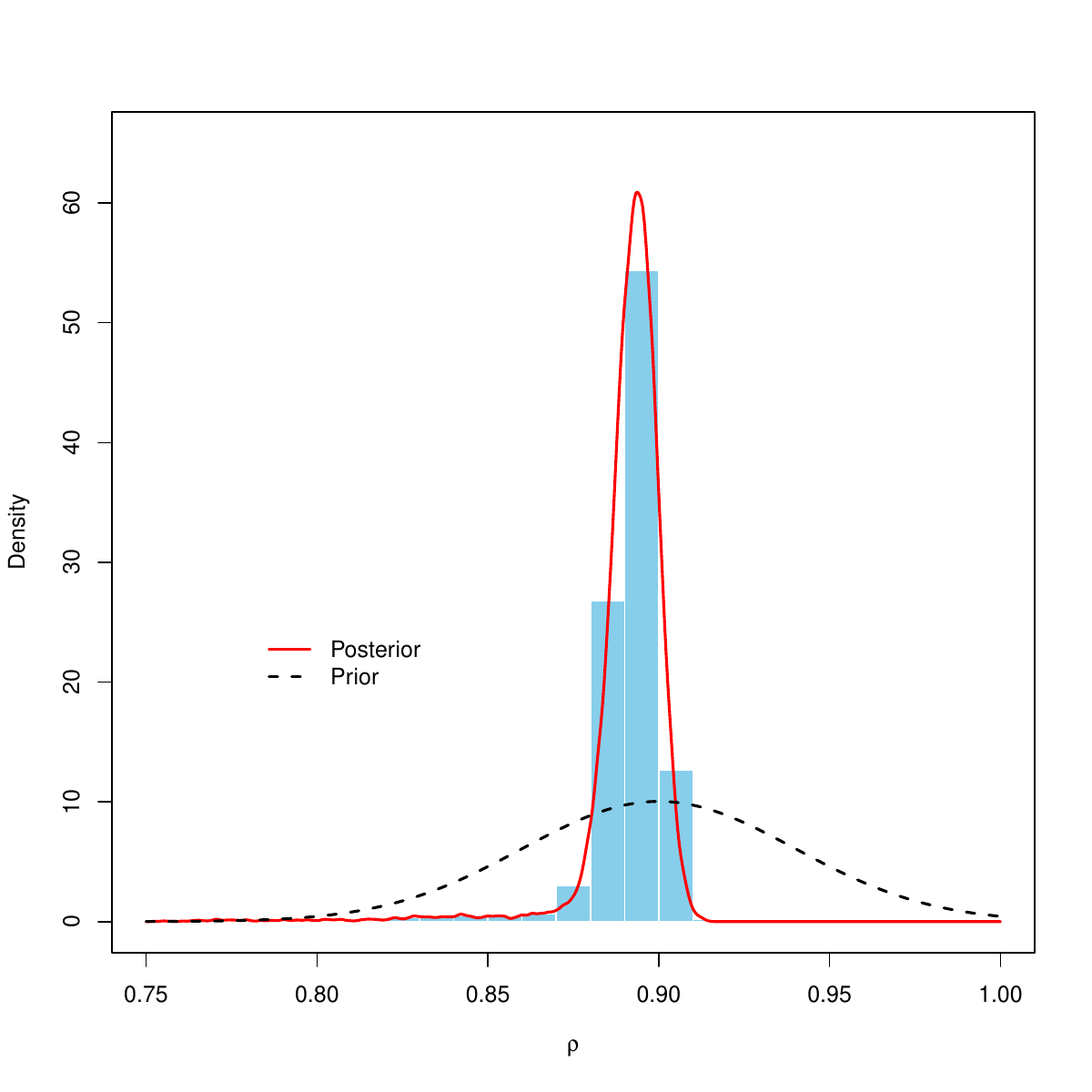}
\caption{\label{fig:prior_vs_post_of_rho} Prior and posterior distributions of the autocorrelation parameter $\rho$.}
\end{figure}

A key hyperparameter in dynamic PUM models is the autocorrelation coefficient $\rho$ of the underlying autoregressive process. To investigate the performance of the model, we might want to compare the prior and posterior distributions for this parameter.   Figure \ref{fig:prior_vs_post_of_rho} shows that, in this dataset, the posterior distribution of $\rho$ is centered close to the prior, but is much more concentrated.  Indeed, its posterior mean is $0.891$, and the associated 95\% posterior credible interval is $(0.863,0.905)$.

\begin{Schunk}
\begin{Sinput}
R> scotus.rho = scotus.pum[["rho"]][["rho"]]
R> xrange = seq(0.75, 1, length.out = 200)
R> hist(scotus.rho, breaks = 25, probability = TRUE, col = "skyblue",
+       border = "white", xlim = c(0.75, 1), ylim = c(0, 65),
+       xlab = expression(rho), ylab = "Density", main = NULL)
R> lines(density(scotus.rho, from = 0.75, to = 1), col = "red", lwd = 2)
R> lines(xrange, dtnorm(xrange, mean = 0.9, sd = 0.04, lower = 0, upper = 1),
+        col = "black", lwd = 2, lty = 2)
R> legend(0.78, 25, legend = c("Posterior", "Prior"),
+         col = c("red", "black"), lty = c(1, 2), lwd = 2, bty = "n")
R> box()
\end{Sinput}
\end{Schunk}

Similarly to the static case, it is illustrative to compare the complexity-adjusted fit of PUM vs.\ that of the more standard dynamic item response theory model proposed by \citet{martin2002dynamic}, which is implemented by the function \code{MCMCdynamicIRT1d} in the package \pkg{MCMCpack} \citep{Martin2011MCMCpack}:
\begin{Schunk}
\begin{Sinput}
R> library(MCMCpack)
R> special_judge_ind = sapply(c("HLBlack", "PStewart", "WHRehnquist"),
+                             function(name){grep(name, rownames(mqVotes))})
R> e0_v = rep(0, nrow(mqVotes))
R> E0_v = rep(1, nrow(mqVotes))
R> e0_v[special_judge_ind] = c(-2, 1, 3)
R> E0_v[special_judge_ind] = c(10, 10, 10)
R> theta.start = rep(0, nrow(mqVotes))
R> indices = c(2, 5, 8, 9, 12, 22, 23, 24, 25, 29, 30, 33, 36, 39,
+              42, 43, 44)
R> values = c(1, 1, -1, -2, -2, 1, -1, 1, 1, -1, 1, 3, 3, 3, 1, 1, -1)
R> theta.start[indices] = values
R> scotus.MQ = MCMCdynamicIRT1d(mqVotes, mqTime, mcmc = 20000,
+    burnin = 30000, thin = 10, verbose = 500, tau2.start = 0.1,
+    theta.start = theta.start, a0 = 0, A0 = 1, b0 = 0, B0 = 1, c0 = -10,
+    d0 = -2, e0 = e0_v, E0 = E0_v,
+    theta.constraints=list(CThomas = "+", SAAlito = "+", WJBrennan = "-",
+                           WODouglas = "-", CEWhittaker = "+"))
\end{Sinput}
\end{Schunk}

Similar to the static case, the functions \code{predict\_pum} and \code{predict\_irt} can be used to calculate the predicted voting probabilities across MCMC samples, and \code{calc\_waic} can be used to evaluate the WAIC score for each term.  The difference between the score under the standard model and PUM can then be displayed as a line plot.
\begin{Schunk}
\begin{Sinput}
R> scotus.pum.predprob = predict_pum(mqVotes, mqTime, scotus.pum)
R> scotus.MQ.predprob = predict_irt(mqVotes, mqTime, scotus.MQ)
R> scotus.pum.waic = calc_waic(mqVotes, mqTime, scotus.pum.predprob)
R> scotus.MQ.waic = calc_waic(mqVotes, mqTime, scotus.MQ.predprob)
\end{Sinput}
\end{Schunk}
\begin{Schunk}
\begin{Sinput}
R> waic_diff = scotus.MQ.waic - scotus.pum.waic
R> years = 1937: 2021
R> plot(years, waic_diff, type = "l", lwd = 2.5, xlab = "Term",
+       ylab = expression(Delta * "WAIC"), ylim = c(-100, 220), cex.lab = 1.5)
R> abline(h = 0, lty = 2, lwd = 2)
R> abline(v = c(1946, 1953, 1969, 1986, 2005), lty = 3)
R> text(c(1949.5, 1961, 1978, 1996, 2015), 160,
+       c("Vinson", "Warren", "Burger", "Rehnquist", "Roberts"),
+       cex = 1.2, pos = 3)
\end{Sinput}
\end{Schunk}

\begin{figure}[t!]
\centering
\includegraphics[width=0.9\textwidth]{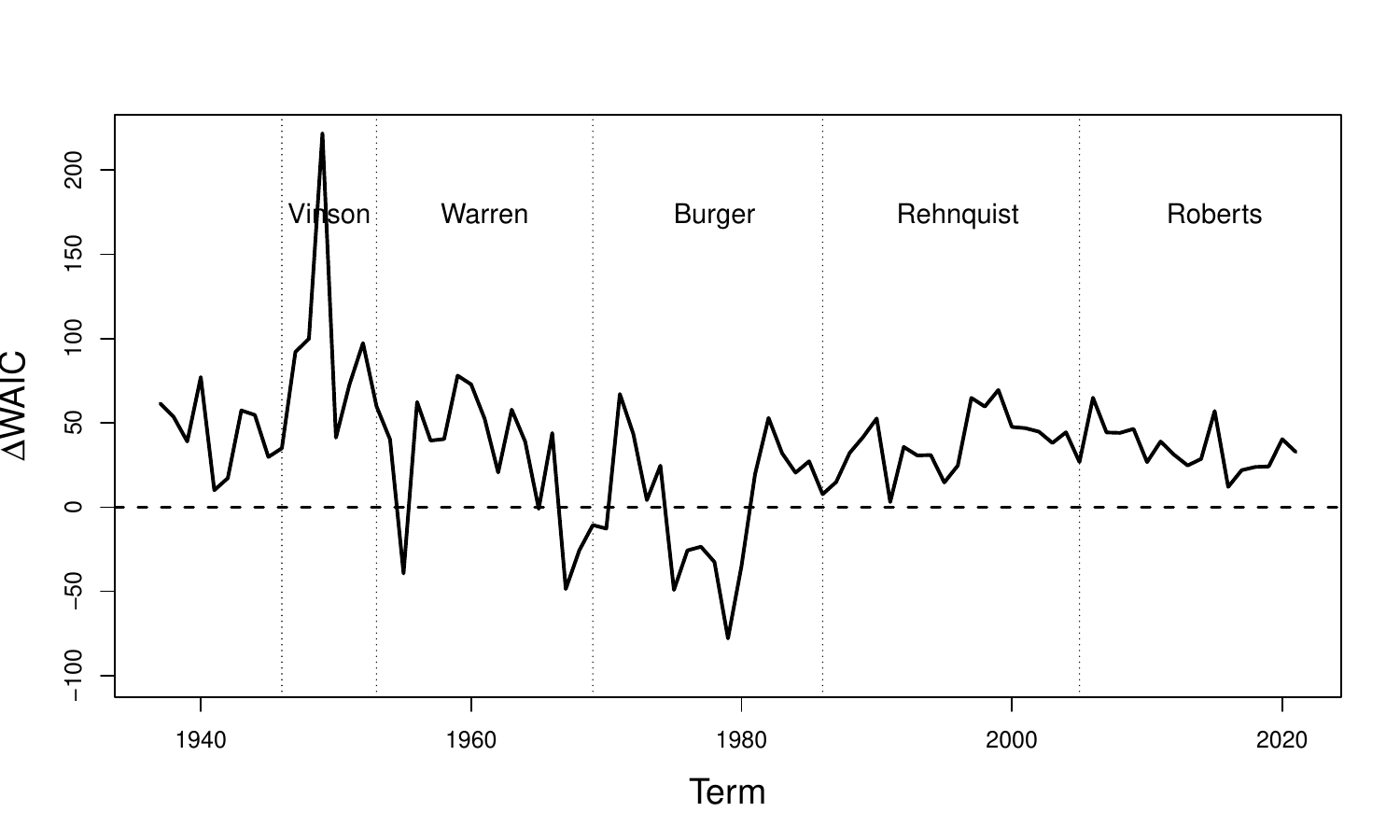}
\caption{\label{fig:plot_waic} WAIC differences between dynamic PUM and IRT models over time.}
\end{figure}

As seen in Figure \ref{fig:plot_waic}, the difference in WAIC is positive on most terms, suggesting that PUM tends to provide a better explanation to the U.S. Supreme Court Justice's voting patterns than the standard dynamic IRT model.

\section{Discussion}\label{sec:discussion}

We have demonstrated the use of \pkg{pumBayes}, an \proglang{R} package that supports Bayesian estimation of static and dynamic probit unfolding models for binary preference data. By allowing non-monotonic response functions, probit unfolding models offer a flexible alternative to traditional item response theory models, especially in political science applications. \pkg{pumBayes}'s flexibility and accuracy make it a valuable tool for researchers analyzing complex voting patterns and ideological behaviors in various political and judicial contexts.  Our hope is that this work will help practitioners in applying this class of models in other contexts.

\bibliography{refs}

\end{document}